\documentclass[a4paper]{article}
\usepackage{amsmath,amssymb}

\begin{document}

\begin{center}
{\Large D1/D5 system and Wilson Loops \\ in (Non-)commutative Gauge Theories}
\end{center}

\vspace{1cm}

\begin{center}
Hidenori TAKAHASHI$^a$ \footnote{htaka@phys.ge.cst.nihon-u.ac.jp},
Tadahito NAKAJIMA$^a$ \footnote{nakajima@phys.ge.cst.nihon-u.ac.jp},

\vspace{5mm}
 $^a$ Laboratory of Physics, College of Science and Technology \\
  Nihon University, Funabashi Chiba 274-8501, Japan

\vspace{5mm}
and

\vspace{5mm}
Kenji SUZUKI$^{b}$ \footnote{ksuzuki@sofia.phys.ocha.ac.jp}

\vspace{3mm}
 $^b$ Department of Physics, Ochanomizu University, \\
 ~~~~Otsuka, Bunkyou-ku, Tokyo 112-8610, Japan
\end{center}


\vspace{2cm}

\begin{center}
{\bf ABSTRACT}
\end{center}

We study the behavior of the Wilson loop in the (5+1)-dimensional supersymmetric Yang-Mills theory with the
presence of the solitonic object. Using the dual string description of the Yang-Mills theory that is given by
the D1/D5 system, we estimate the Wilson loops  both in the temporal and spatial cases.
  For the case of the temporal loop, we obtain the velocity dependent potential.
For the spatial loop, we find that the area law is emerged due to the effect of the D1-branes.
Further, we consider D1/D5 system in the presence of the constant $B$ field. It is found that the Wilson loop
 obeys the area law for the effect of the noncommutativity.

\newpage

%
%
%
%
%
%
%
%
%
%
%

\section{Introduction}
The large $N$ limit of the conformal field theory can be described by the supergravity
 on $AdS \times {\rm(manifold)}$, which is known as  AdS/CFT correspondence
\cite{malda_ATMP2,witten_ATMP2,gub_kle_pol,ah_gu_ma_oo_oz}.
For example, $\mathcal{N}$=4 supersymmetric Yang-Mills theory
in (3+1)-dimensions can be described by type IIB string theory on $AdS_5 \times S^5$ with the appropriate boundary
 conditions. The supersymmetry group of $AdS_5 \times S^5$ is equivalent to the superconformal group
 in (3+1)-dimensions.
The radius $R$ over string length $\ell_s$ of the $AdS_5$ and $S^5$ is given by 
$R^4/\ell_s^4 \sim gN \sim g_{\rm YM}^2N$. The supergravity description is valid in the $R \gg \ell_s$ region,
which correspond to taking the  't Hooft limit of the Yang-Mills theory. Therefore, the strong coupling limit
of the Yang-Mills theory can be described by the low energy limit of the superstring theory.
Using this duality, we can estimate the quark-antiquark potential for the ${\mathcal N}=4$ supersymmetric Yang-Mills
theory. Maldacena \cite{malda_PRL80}, Rey and Yee \cite{rey_yee}  show that the potential is given by the Coulomb type,
\begin{equation}
E_{q \bar q} =  - \frac{4 \pi^2 R^2}{\Gamma(1/4)^4} \frac{1}{L}.
\end{equation}
Furthermore, the $N$ coincident extremal D$p$-branes system can describe the $(p+1)$-dimensional $U(N)$ supersymmetric
Yang-Mills theory with sixteen supercharges \cite{itzh_mald_sonn_yank} in the decoupling limit.

Similarly, the large $N$ limit of the noncommutative gauge theories are described by
 the D$p$-branes system with a constant $B$ field in the near horizon region \cite{malda_russ,has_itz,alis_oz_shei}.
It is well-known that the noncommutative Yang-Mills theory is equivalent to ordinary Yang-Mills theory
with the perturbative higher dimension operator \cite{sei_wit}. At the long distance (IR regime),
the dual gravity description of the noncommutative supersymmetric Yang-Mills theories becomes the $AdS_5 \times S^5$.
Then, the noncommutative gauge theories reduce to the commutative ones. 
However, close to the boundary (UV regime) with taking $B$ field to infinity, the noncommutativity effect
 becomes strong. 
Therefore, the dual gravity solutions behave differently from the commutative ones.
 We can see this noncommutativity effects in the Wilson loop as shown in \cite{dhar_kita}.
For the large loop, that corresponds to the IR limit with small noncommutativity effects,
  the Wilson loop shows a Coulomb law behavior \cite{malda_russ,alis_oz_shei}.
On the other hand, for the small loop, we find the area law and the string tension is controlled by the
noncommutativity scale \cite{dhar_kita,lee_sin}.

In type IIB string theory, there is the black hole solution with non zero horizon area, which is known as 
D1/D5 system \cite{stro_vafa}.
In the context of the AdS/CFT correspondence, the D1/D5 system whose near horizon geometry is
$AdS_3 \times S^3 \times M^4$ can describe the (1+1)-dimensional conformal field theory
having eight supersymmetries \cite{malda_ATMP2}.
One possible interpretation of the CFT is that the target space of the CFT is the moduli space of
 the instantons \cite{doug,dijk}.
Small fluctuations of the instanton moduli are described by a (1+1)-dimensional sigma model.
The extension to the noncommutative space-time is given by \cite{dijk,dha_man_wad_yog,mikh}. In this case,
the background metric of the string theory is deformed to $AdS_3 \times S^3 \times X$ where $X$ is the four dimensional
manifold. For the effect of $B$ field, $X$ become noncommutative manifold \cite{mikh}.

More generally, we can expect that the non-conformal gauge theories can be described by the corresponding
gravity theories \cite{ah_gu_ma_oo_oz}, e.g. the D-instanton + D3-brane configuration, shortly denoted by
 the D(-1)/D3 system \cite{liu_tse}, which is $T$-dual to the D1/D5 system. The D3 branes system 
is dual to the CFT and the quark-antiquark potential is Coulombic. However, the D(-1)/D3 system is dual to
the nonconformal theory and the potential is linear in $L$ due to the existence of the D-instantons.
For the case of the D5 branes system, it is dual to the super Yang-Mills \cite{itzh_mald_sonn_yank} and 
the Wilson loop has  a zero value \cite{bra_itz_son_yan}. How does the physics of the D5 branes system change
if we introduce the D1-branes?

In this letter, we consider the D1/D5 system which is  the dual string description of the (5+1)-dimensional gauge
theory with the solitonic object \cite{mald_nps98,dav_man_wad}.
In this case, D1-branes can be interpreted as the instantons \cite{doug}. 
We estimate the Wilson loop of the gauge theory using the dual string description.
By taking the appropriate decoupling limit, we find the quark-antiquark potential. For the temporal loop, 
we have the velocity dependent potential. On the other hand, the spatial loop shows the area law
 for the effect of the D1-branes.
Further, we consider the D1/D5 system on the noncommutative space and estimate the Wilson loop.
In this case, we have a area law even when the noncommutativity becomes strong.

\section{Temporal Wilson Loop in the moving frame}
The supergravity solution corresponding to the extremal D1/D5 system is given by \cite{malda_ATMP2}
\begin{equation}
ds^2 = f^{-1/2}_1 f^{-1/2}_5 ( -d t^2 + d x^2_1)
    + f^{1/2}_1 f^{-1/2}_5 ( d x^2_2 +  \cdots + d x^2_5 )
    + f^{1/2}_1 f^{1/2}_5 (d r^2 + r^2 d \Omega^2_3 ),
\end{equation}
where
\begin{equation}
f_1 = 1+\frac{R_1^2}{U^2}, \quad \quad
f_5 = 1+\frac{R_5^2}{\alpha^{\prime \,2} U^2}.
\label{def_f1_f5}
\end{equation}
Here, we define
\begin{equation}
R_1^2 \equiv \frac{(2 \pi)^4 \alpha^\prime g Q_1}{V_4}, \quad
R_5^2 \equiv \alpha^\prime g Q_5, \quad U=\frac{r}{\alpha^\prime},
\end{equation}
and $V_4$ is the volume of $M_4$. We take the decoupling limit with
keeping the volume of the $M_4$ finite as
\begin{equation}
 \begin{split}
  &\alpha^\prime \rightarrow 0, \quad r/\alpha^\prime = \textrm{fixed}, \quad
   V_4/(2\pi)^4 = \textrm{fixed},  \\
  &R_1^2 = \textrm{fixed}, \quad R_5^2 = \textrm{fixed}.
\end{split}
\label{decup_lim_com}
\end{equation}
In this limit, the metric becomes
\begin{equation}
ds^2/\alpha^\prime = \frac{U}{R_5 f^{1/2}_1}( -d t^2 + d x^2_1)
    + \frac{U f^{1/2}_1}{R_5} ( d x^2_2 + \cdots + d x^2_5 ) 
    + U R_5 f^{1/2}_1 \left(\frac{d U^2}{U^2} + d \Omega^2_3 \right).
\end{equation}
Now, we calculate the quark-antiquark potential. The Wilson loop can be regarded as the phase factor of a
heavy massive quark \cite{malda_PRL80}. We consider a rectangular loop with sides $T$ and $L$.
Here, $T$ and $L$ are identified with time direction and the distance between quark and antiquark, respectively.
The Wilson loop averages are postulated to be given by minimizing the relevant Nambu--Goto (NG) action, 
\begin{eqnarray}
S=\frac{1}{2\pi\alpha'}\int d\tau d\sigma \sqrt{\det (G_{MN}\partial_\mu X^M \partial_\nu X^N)} \, .
\end{eqnarray}
We will consider the moving string with velocity $v$ in $X^3$ direction. Then,
we take the configuration as
\begin{equation}
X^0=\tau, \quad X^2= \sigma, \quad X^3=v \tau, \quad U=U(\sigma).
\label{temp_conf}
\end{equation}
In this case, the NG action  becomes
\begin{equation}
S = \frac{T R_1R_5}{ 2 \pi} \int \! d \sigma \;
    \sqrt{(1-v^2 f_1) \left( u^{\prime \, 2}+ H u^4 \right)},
\end{equation}
where $u^\prime = \partial_\sigma u$ and
\begin{equation}
H \equiv \dfrac{R_1^2}{U^2} = \dfrac{1}{R_5^2 u^2},
\end{equation}
and we rescale the parameter as
\begin{equation}
u=\frac{U}{R_1 R_5}.
\end{equation}
 At extremum of the action, we find
\begin{equation}
H u^4 \sqrt{\frac{1-v^2 f_1}{u^{\prime \, 2}+ H u^4}} = \text{const.}
\end{equation}
We take  $u=u_0$ at $\partial_\sigma u = 0$, then we have
\begin{equation}
\partial_\sigma u = C^{-1} \sqrt{H_0} u \sqrt{u^2-u_0^2},
\end{equation}
where $H_0=H(u_0)$ and the constant $C$ is given by
\begin{equation}
C = \dfrac{\sqrt{1-v^2 f_1(u_0)}}{\sqrt{1-v^2}}.
\end{equation}
Therefore, the $u_0$ is determined by the condition
\begin{equation}
\frac{L}{2} = \frac{C}{u_0 \sqrt{H_0}} \int_{1}^{\infty} \! \frac{dy}{y \sqrt{y^2-1}}
            =  \frac{\pi C R_5}{2},
\end{equation}
where $y=u/u_0$. Note that it is consistent with \cite{bra_itz_son_yan} when $v=0$.
Now, we compute the energy of the quark-antiquark pair. The energy is given by
\begin{equation}
E = \frac{S}{T} =\frac{R_1R_5 u_0}{\pi \sqrt{1-v^2}} \int \! dy \;\left[
    \frac{(1-v^2)y}{\sqrt{y^2-1}} - \frac{v^2 H_0}{y\sqrt{y^2-1}} \right].
\end{equation}
We subtract the infinite mass of the two quarks which corresponds to the energy of the stretching strings
 \cite{ah_gu_ma_oo_oz,malda_PRL80}. Therefore, the energy is given by
\begin{equation}
E = E(u_0)-E(u_0=0)  = - \dfrac{v^2 R_1}{2 \sqrt{1-v^2}} \sqrt{H_0}.
\end{equation}
Finally, we have
\begin{equation}
E = - \dfrac{v R_1}{2} \sqrt{1-\dfrac{8 \pi }{g_{YM}^2 Q_5}L^2},
\label{vel_potential}
\end{equation}
where the coupling constant of the Yang-Mills theory is given by
$g_{YM}^2 = (2 \pi)^3 g \alpha^\prime$ \cite{itzh_mald_sonn_yank} on $p=5$.
The potential depends on the speed of the moving frame and the D1 charge.
More generally, when we consider the moving string with the velocity $\mbox{\boldmath $v$}=(v_3, v_4, v_5)$
in $X^3, X^4, X^5$ directions, the potential is proportional to the speed $v=\sqrt{v_3^2+v_4^2+v_5^2}$.
The ``instantons" affect the system and the attractive force is emerged. We also note that 
the distance $L$ does not exceed the $g_{YM} Q_5^{1/2}/(8 \pi)^{1/2}$. 
The maximum of the distance $L$ corresponds to $u_0=\infty$. At this point the energy becomes zero.

\section{Spatial Wilson Loop in the noncommutative space}
Next, we consider the D1/D5 system in the presence of the constant $B$ field.
Here, we take the nonzero component of the $B$ field at $x_2,x_3$ directions. In this case,
the solution with the Euclidean signature is given by \cite{malda_russ}
\begin{eqnarray}
ds^2 &=& f^{-1/2}_1 f^{-1/2}_5 ( d t^2 + d x^2_1) 
    + f^{1/2}_1 f^{-1/2}_5 \left[ \tilde{h}(d \tilde{x}^2_2 +d \tilde{x}^2_3)+ d x^2_4 + d x^2_5 \right]
\nonumber\\
    && \hspace{5mm}+ f^{1/2}_1 f^{1/2}_5 (d r^2 + r^2 d \Omega^2_3 ), 
\end{eqnarray}
where $f_1, f_5$ is defined in eq.(\ref{def_f1_f5}) and
\begin{equation}
\tilde h^{-1} = (f_1/f_5)\sin^2\theta+\cos^2\theta.
\end{equation}
We take the decoupling limit of eq.(\ref{decup_lim_com}) with
\begin{equation}
b= \alpha'\tan \theta \equiv a R_5, \quad \tilde{x}_2 = \frac{\alpha'}{b} x_2, \quad
\tilde{x}_3 = \frac{\alpha'}{b} x_3.
\end{equation}
In this case, the metric becomes
\begin{eqnarray}
ds^2/\alpha^\prime &= \dfrac{U}{R_5\sqrt{f_1}} (- d t^2 + d x^2_1)  + \dfrac{U}{R_5}\sqrt{f_1} 
  \left[h(d x^2_2 +  d x^2_3)+ d x^2_4+ dx^2_5 \right]  \nonumber \\
  &+ U R_5\sqrt{f_1}\left(\dfrac{dU^2}{U^2} + d \Omega^2_3 \right),
\end{eqnarray}
where $h^{-1}=1+a^2 U^2 f_1$.
Since we now consider the spatial loop,  we take the configuration as,
\begin{eqnarray*}
X^4 = \tau, \quad X^2 = \sigma, \quad U = U(\sigma).
\end{eqnarray*}
In this case, the Nambu--Goto action becomes
\begin{equation}
S = \frac{T}{ 2 \pi} \int \! d \sigma \;
    \sqrt{f_1 \bigg(U^{\prime \, 2} + h\frac{U^2}{R_5^2} \bigg)} .
\end{equation}
Similarly, defining $U_0$ with $\partial_\sigma U|_{U=U_0}=0$, we have
\begin{equation}
\partial_\sigma U = \frac{hU}{U_0 R_5\sqrt{f_1^{(0)}}} \sqrt{U^2-U_0^2},
\end{equation}
where $f_1^{(0)}=f_1(U_0)$. Finally, the distance of the quark-antiquark $L$ is given by
\begin{equation}
\frac{L}{2} = R_5 \sqrt{f_1^0}\int_1^\infty \!dy \left[
\frac{1+a^2R_1^2}{y\sqrt{y^2-1}} + \frac{(a U_0)^2 y}{\sqrt{y^2-1}} \right],
\end{equation}
where $y=U/U_0$. For the noncommutaitve case, the distance $L$ becomes infinite at $aU_0 \gg 1$ (UV limit)
 \cite{dhar_kita,lee_sin}. This can be occurred as the effect of the noncommutativity.
In the present case, since the second term which has the noncommutative parameter $a$ is dominant at the UV regime.
the distance becomes infinite. Dhar and Kitazawa \cite{dhar_kita} discuss the system of D3 branes
in the presence of $B_{23}$ field and suggest that the boundary is located at a finite value of $U$.
Therefore, for the case of the D1/D5 system with $B_{23}$ field, the boundary might also be located at $U=\Lambda$. 

The energy of the quark-antiquark pair is given by
\begin{equation}
E = \frac{U_0}{\pi \sqrt{h_0}}\int_1^{\Lambda/U_0}\!\!dy \left[
\frac{R_1^2/U_0^2}{y\sqrt{y^2-1}} + \frac{y}{\sqrt{y^2-1}} \right],
\end{equation}
where $h_0^{-1}=1+a^2 U_0^2 f_1^{(0)}$. When the noncommutativity is large ($aU_0 \gg 1$), we obtain
\begin{equation}
\frac{L}{2} \simeq R_5 a^2 U_0 \sqrt{\Lambda^2-U_0^2}, \quad E \simeq \frac{a U_0}{\pi} \sqrt{\Lambda^2-U_0^2}.
\end{equation}
Therefore, we get the linear potential
\begin{equation}
E = \frac{1}{2\pi aR_5}L = \frac{\sqrt{2\pi}}{a\sqrt{Q_5} g_{YM}}L,
\label{strong_non_e}
\end{equation}
where $g_{YM}$ is the coupling constant of the Yang-Mills theory in the (5+1)-dimensions.
Note that the string tension depends on the inverse of the noncommutative parameter $a$ and the Yang-Mills coupling.
Therefore, the area law behavior could be caused by the nonperturbative effect.
On the other hand, when D1 charge $Q_1$ is large ($R_1 \gg 1$) and the noncommutativity effect is small ($aU_0 \ll 1$),
the first term of $L$ and $E$ is dominant. Then, D1 branes play an important role in this region.
In this case, $L$ and $E$ are given by
\begin{equation}
\frac{L}{2} \simeq \frac{R_1 R_5}{U_0}(1+ a^2 R_1^2) g(\Lambda), \quad
  E \simeq \frac{R_1^2}{ \pi U_0} \sqrt{1+ a^2 R_1^2} g(\Lambda),
\end{equation}
where 
\begin{equation}
g(\Lambda)= \tan^{-1} \frac{\sqrt{\Lambda^2-U_0^2}}{U_0}.
\end{equation}
Therefore, we get the linear potential as
\begin{equation}
 E =\frac{R_1}{2 \pi R_5\sqrt{1+a^2R_1^2}} L.
\end{equation}
Although, in the commutative space-time ($a=0$), the potential of the quark-antiquark is linear in $L$;
\begin{eqnarray}
 E = \frac{R_1}{2 \pi R_5}L.
\end{eqnarray}
Note that in the D3 brane system, the potential becomes the Coulombic when the noncommutativity effect is small
\cite{dhar_kita}. However, in D1/D5 system, the potential is linear in $L$.

\section{Summary and Concluding remarks}
We have considered the (5+1)-dimensional gauge theory in the presence of the solitonic object
 as the low energy theory of the D1/D5 system. Using this dual description, 
we can estimate the Wilson loop in the (non)-commutative space along the same strategy of
\cite{malda_PRL80,rey_yee}. However, we take the decoupling limit with keeping the volume of the $M^4$ finite.
Further, for the noncommutative space, the constant $B$ field is introduced in the supergravity theory. 
Thus, the part of the supersymmetry is broken in the noncommutative Yang-Mills theory.
Therefore, the Yang-Mills theory becomes the nonconformal theory.

In the temporal loop, we have the velocity dependent potential. In the D5 system, the distance $L$ is independent
of $U_0$ and the energy is zero \cite{alis_oz_shei,bra_itz_son_yan}.
However, in the D1/D5 system, the potential becomes nontrivial due to the effect of the D1 branes.
This may be the ``instantons" effect.
The potential eq.(\ref{vel_potential}) implies that the distance of the quark-antiquark must satisfy
\begin{equation}
L < (8\pi)^{-1/2} g_{YM} \sqrt{Q_5}.
\end{equation}
This relation was found in \cite{bra_itz_son_yan}, which must be related to the existence
 of a non-locality scale \cite{bra_itz_son_yan}.

In the spatial loop, which corresponds to the Euclidean field theory, we have the area law 
 both in the commutative and the noncommutative space. Then, the potential of the quark and antiquark behaves linearly.
In UV region ($aU_0 \gg 1$), the noncommutativity effect is crucial and the string tension is controlled by the 
noncommutativity parameter. 
 Note that, in our case,  the string tension depends on $1/g_{YM}$. Therefore the result suggests
that it is the nonperturbative phenomena.  In noncommutative gauge theories, the area law behavior could be
occurred by the nonperturbative effect. Unfortunately, at the present, we do not know how 
the nonperturbative effect can be described in the noncommutative gauge theory view point.
But, the area law behavior of the noncommutative Yang-Mills theory in the strong commutativity can be universal
\cite{lee_sin}, and then it could be caused by the nonperturbative effect.

It is interesting to note that even in IR region ($aU_0 \ll 1$) where the noncommutativity effect is small,
we also have the area law. This is the different point against the D3 brane case \cite{dhar_kita}. 
However, Liu and Tseytlin \cite{liu_tse} consider $SO(4) \times SO(6)$ invariant type IIB string solution
describing D3 branes with D-instantons distributed over D3 brane world volume, which is known as the D(-1)/D3 system.
 They estimate the Wilson loop and show that the area law is emerged. The string tension of D(-1)/D3 system
 depends on the D-instanton density. Since the D-instanton is related to the self-dual gauge field,
 the gauge field condensation leads to the area law behavior of the Wilson loop \cite{liu_tse}.
Similarly, in the D1/D5 system, the Wilson loop shows the area law and the string tension is
 proportional to the D1 charge. The D1 brane is interpreted as the instanton carrying
 the D1 charge $Q_1$ \cite{mald_nps98}. Therefore, the Wilson loop satisfies the area law for
the effect of the ``instantons".
 We also note that the D1/D5 system is $T$-dual to D(-1)/D3 or D0/D4 system. 
 Then, the present result is consistent with Liu and Tseytlin \cite{liu_tse}.
Finally, the confinement should be caused by these solitonic object \cite{liu_tse}.

In several cases of the near extremal solutions or the finite temperature theories
\cite{witten_ATMP2_505,bra_itz_son_yan,ksuzuki}, we have the area law behavior of the Wilson loop.
The non-zero temperature breaks supersymmetry and the  conformal invariance.
 Therefore, we can expect that even for the non-supersymmetric case,
the strongly coupled gauge theories can be described by the dual string theories.
In this letter, we consider the {\it extremal} branes solution and find the area law behavior of the Wilson loop.
In this case, the area law behavior is caused by the effect of the D1 branes and the $B$ field.
Further, the part of the supersymmetry can be broken in our limiting procedure. 
Therefore, the corresponding gauge theory is the non-conformal theory.

It would be interesting to study whether the area law behavior of the Wilson 
loop is universal in other (nonconformal) gauge theories 
by using the dual description of more general $Dp/Dp'$ systems (with constant B-field).

\vspace{1.5cm}
{\bf \Large Acknowledgements}

\vspace{5mm}
We would like to thank T. Asaga and T. Fujita  for reading manuscript and comments.

\end{document}